\documentclass[a4paper,aps,twocolumn]{revtex4}
\RequirePackage[english]{babel}
\RequirePackage[latin1]{inputenc}
\RequirePackage[T1]{fontenc}
\RequirePackage{mathrsfs}
\RequirePackage{amsmath}
\RequirePackage{amssymb}
\RequirePackage{amsbsy}
\RequirePackage{bm}
\begin{document}
\title{\bf{Torsionally gravitating charged matter fields and quanta}}
\author{Luca Fabbri}
\affiliation{INFN \& Dipartimento di Fisica, Universit\`{a} di Bologna, Via Irnerio 46, 40126 Bologna, ITALY}
\date{\today}
\begin{abstract}
In the present article we shall consider the torsional completion of a gravitational background that is filled with electrodynamically interacting material fields, taken to be of fermionic type, eventually deriving properties like the impossibility of singularities and the possibility of confinement, both necessary for a correct quantum description.
\end{abstract}
\maketitle
\section{INTRODUCTION}
In the theory of classical fields the natural interpretation given in terms of real distributions is made difficult by many factors, and among them what for us is the most important is that the inevitable spreading of matter would not fit with the natural localization of particles.

This problem seems to have no solution because for electrically charged matter the electrodynamic interaction of the field with itself is repulsive and stronger than the self-gravitating attraction; swamping the only source of pull vanishes the only possibility of confinement unless new physics is considered, and luckily the gravitational background is already endowed with additional interactions if the torsion of the spacetime is not neglected.

When the spacetime torsion is allowed to remain the underlying gravitational field is modified in such a way that at short distances it is no longer negligible with respect to the electrodynamic field thus providing the conditions for the material confinement; the new tendency to localize in compact regions would now give rise to the inverse problem of singularity formation, but luckily again this is avoided by the fact that the matter field has an effective repulsion due to torsion: then the presence of torsion has the consequence of both modifying the gravitational field with an additional attraction and giving rise within the matter field equations to an effective repulsion in such a way that a fermionic matter distribution would be confined in a localized region while being not singular and such equilibrium would also be stable as required for the quantum systems in the common paradigm.
\section{Torsionally Gravitating Charged Matter Fields}
To build the foundation of the field equations we shall employ we consider one hypothesis, that the dimension of the underlying background space as well as the dimension of the spin representation and the order derivative of the field equations be fixed at their least value allowed.

This sort of principle of simplicity might be justified by the observation that, if we were to increase the number of variables or the number of independent components of a field or again the order of derivation in the field equations, then any of these actions would result into an increase of the number of integrations, thus the number of adjustable parameters, losing predictivity: if we want that our construction be the most stringent then the dimension of the space and the spin and the order derivative of the field equations have to be taken at their least value, and this means that we have to take $(1\!+\!3)$-dimensional spacetimes filled with $\frac{1}{2}$-spin spinor fields and all field equations are assigned at the lowest-order derivative permitted.

So after that this specification is implemented, we can see that the Cartan-Riemann geometry \cite{C1,C2,C3,C4} is achieved by introducing the torsion and metric tensors, or equivalently the torsion and curvature tensors, and the following Sciama-Kibble-Einstein theory \cite{S,K} is accomplished by assigning field equations coupling torsion to the spin and curvature to the energy \cite{h-h-k-n}; because torsion is a tensor and the torsion-spin coupling is algebraic, then it is possible to have all geometrical entities decomposed in terms of the correspondent torsionless geometrical entities plus torsional contributions written in terms of the spin of matter, and the torsional completion of gravitation can be shown to be equivalent in its final form to the simplest torsionless gravitation supplemented with torsionally induced spin-contact interactions of a very specific structure \cite{Fabbri:2013isa,Fabbri:2014zya,Fabbri:2014dxa}: in the above references there is the complete development, here we present the final form.

As for the content, the metric $g_{\mu\nu}\!=\!\eta_{ab}e^{a}_{\mu}e^{b}_{\nu}$ is given in terms of the Minkowskian metric and the basis of tetrad fields with which one can compute the connection and the spin-connection $\Lambda^{\alpha}_{\mu\nu}$ and $\Lambda^{i}_{\phantom{i}j\nu}$ and therefore the curvature tensor $R_{\rho\sigma\mu\nu}$ called Riemann curvature tensor, with contractions $R^{\rho}_{\phantom{\rho}\mu\rho\nu}\!\!=\!\!R_{\mu\nu}$ and $R_{\eta\nu}g^{\eta\nu}\!\!=\!\!R$ which are called Ricci curvature tensor and scalar; from the gauge potential indicated by $A_{\mu}$ it is possible to calculate the gauge strength $F_{\mu\nu}$ called Maxwell strength: Clifford matrices are expressed as $\boldsymbol{\gamma}_{\alpha}$ with $4\boldsymbol{\sigma}_{ab}\!=\![\boldsymbol{\gamma}_{a},\boldsymbol{\gamma}_{b}]$ and such that they verify $\{\boldsymbol{\sigma}_{ab},\boldsymbol{\gamma}_{c}\}\!=\!i\varepsilon_{abcd}
\boldsymbol{\pi}\boldsymbol{\gamma}^{d}$ allowing for the definition of the spinor fields $\psi$ and $\overline{\psi}$ and the spinorial covariant derivatives $\boldsymbol{\nabla}_{\mu}$ acting on the spinor fields themselves.

It is then possible to write down what according to our initial hypotheses is the most general Lagrangian as
\begin{eqnarray}
\nonumber
&L=R\!+\!\frac{1}{4}F^{\mu\nu}F_{\mu\nu}
\!-\!\frac{i}{2}(\overline{\psi}\boldsymbol{\gamma}^{\mu}\boldsymbol{\nabla}_{\mu}\psi
\!-\!\boldsymbol{\nabla}_{\mu}\overline{\psi}\boldsymbol{\gamma}^{\mu}\psi)+\\
&+\frac{1}{2}Y\overline{\psi}\boldsymbol{\gamma}_{\mu}\psi
\overline{\psi}\boldsymbol{\gamma}^{\mu}\psi
\!+\!m\overline{\psi}\psi
\end{eqnarray}
determining the dynamics of the system we will study.

By varying this Lagrangian we get the field equations given by the torsionless symmetric curvature-energy coupling field equations describing the gravitational field and the strength-current coupling field equations describing the electrodynamic field together with the spinorial field equations describing the dynamics of the matter field as
\begin{eqnarray}
\nonumber
&(R^{\rho\alpha}\!-\!\frac{1}{2}Rg^{\rho\alpha})
\!+\!\frac{1}{2}(F^{\rho\sigma}F^{\alpha}_{\phantom{\alpha}\sigma}
\!-\!\frac{1}{4}g^{\rho\alpha}F^{2})=\\
\nonumber
&=\frac{i}{8}(\overline{\psi}\boldsymbol{\gamma}^{\rho}\boldsymbol{\nabla}^{\alpha}\psi
\!-\!\boldsymbol{\nabla}^{\alpha}\overline{\psi}\boldsymbol{\gamma}^{\rho}\psi+\\
&+\overline{\psi}\boldsymbol{\gamma}^{\alpha}\boldsymbol{\nabla}^{\rho}\psi
\!-\!\boldsymbol{\nabla}^{\rho}\overline{\psi}\boldsymbol{\gamma}^{\alpha}\psi)
\!-\!\frac{1}{4}Y\overline{\psi}\boldsymbol{\gamma}_{\mu}\psi
\overline{\psi}\boldsymbol{\gamma}^{\mu}\psi g^{\alpha\rho}
\label{gravitational}\\
&\nabla_{\sigma}F^{\sigma\rho}\!=\!q\overline{\psi}\boldsymbol{\gamma}^{\rho}\psi
\label{electrodynamical}\\
&i\boldsymbol{\gamma}^{\mu}\boldsymbol{\nabla}_{\mu}\psi
\!-\!Y\overline{\psi}\boldsymbol{\gamma}_{\rho}\psi
\boldsymbol{\gamma}^{\rho}\psi\!-\!m\psi\!=\!0
\label{material}
\end{eqnarray}
in which we recognize the presence of the additional non-linear potentials: if we proceed by contracting the gravitational field equation (\ref{gravitational}) then it is possible to see that this system of field equations has the equivalent form
\begin{eqnarray}
\nonumber
&R^{\rho\alpha}\!+\!\frac{1}{2}(F^{\rho\sigma}F^{\alpha}_{\phantom{\alpha}\sigma}
\!-\!\frac{1}{4}g^{\rho\alpha}F^{2})=\\
\nonumber
&=\frac{i}{8}(\overline{\psi}\boldsymbol{\gamma}^{\rho}\boldsymbol{\nabla}^{\alpha}\psi
\!-\!\boldsymbol{\nabla}^{\alpha}\overline{\psi}\boldsymbol{\gamma}^{\rho}\psi+\\
&+\overline{\psi}\boldsymbol{\gamma}^{\alpha}\boldsymbol{\nabla}^{\rho}\psi
\!-\!\boldsymbol{\nabla}^{\rho}\overline{\psi}\boldsymbol{\gamma}^{\alpha}\psi)
\!-\!\frac{1}{4}m\overline{\psi}\psi g^{\alpha\rho}
\label{gravitation}\\
&\nabla_{\sigma}F^{\sigma\rho}\!=\!q\overline{\psi}\boldsymbol{\gamma}^{\rho}\psi
\label{electrodynamics}\\
&i\boldsymbol{\gamma}^{\mu}\boldsymbol{\nabla}_{\mu}\psi
\!-\!Y\overline{\psi}\boldsymbol{\gamma}_{\rho}\psi
\boldsymbol{\gamma}^{\rho}\psi\!-\!m\psi\!=\!0
\label{matter}
\end{eqnarray}
making clear that only within the matter field equations there are torsionally induced spin contact interactions described by non-linear potentials yielding effective forces.

In the stationary case the low-speed weak-gravity limit is given when the metric has time-time component that is given according to $g_{tt}\!\approx\!1\!+\!2u$ in terms of the gravitational potential from which it is possible to calculate the gravitational energy indicated by $U$ and the gauge potential has time component given by $A_{t}\!\approx\!w$ in terms of the electrostatic potential from which it is possible to compute the electrostatic energy indicated with $W$ so that by considering the spinor energy $E$ we have that in standard representation the spinor field is $\overline{\psi}\!\approx\!(\phi^{\dagger},0)$ and thus
\begin{eqnarray}
&\mathrm{div}\ \mathrm{grad}u\!\approx\!\frac{1}{4}|\mathrm{grad}w|^{2}
\!+\!\frac{1}{2}Y|\phi^{\dagger}\phi|^{2}\!+\!\frac{1}{4}m\phi^{\dagger}\phi
\label{gravity}\\
&\mathrm{div}\ \mathrm{grad}w\!\approx\!-q\phi^{\dagger}\phi
\label{electrostatic}\\
&\frac{1}{2m}\!\vec{\boldsymbol{\nabla}}\!\cdot\!\vec{\boldsymbol{\nabla}}\phi
\!-\!(\frac{Y^{2}}{2m}|\phi^{\dagger}\!\phi|^{2}\!\!+\!\!Y\!\phi^{\dagger}\!\phi
\!+\!U\!\!+\!W)\phi\!+\!\!E\phi\!\approx\!0
\label{mass}
\end{eqnarray}
as the field equations we will investigate in the following.

The constant $Y$ is undetermined and $m$ and $q$ are the mass and charge of the particle associated to the matter.
\section{Stably Equilibrated Distributions}
After this introduction of the foundations, we start to analyze the consequences, and one initial thing we point out is that the torsional constant $Y$ has the same dimension of the gravitational Newton constant but in general it does not need to have the same value, and in fact it does not even need to have the same sign; indeed, the reversal of the sign of this constant is a trick that has been used in different occasions in order to discuss some problems connected to singularity formation and confinement.

For example, recently in \cite{Magueijo:2012ug} it has been demonstrated that in field equations (\ref{gravitation}) a negative $Y$ has the effect that at high densities the energy becomes negative, so that the gravitational field turns out to be repulsive and a gravitational singularity may be avoided; analogously in the seminal paper \cite{n-j--l} it has been discussed how within the field equations (\ref{matter}) a negative $Y$ turns the non-linear potential into an attractive potential, so that when this attractive potential is balanced by the spreading tendency of the massive field a condition of equilibrium may be reached, as mechanism of condensation: in the former approach the authors insist on the gravitational sector while in the latter work the authors deal only with the matter field equations, but in both cases the idea is that when the constant $Y$ is negative then the non-linear terms and the mass term affect the dynamics in opposite ways, and a condition for the equilibrium may arise eventually.

On the other hand, however, such an equilibrium would not be stable under external perturbations: if an external field is applied in such a way as to increase the density of the distribution, then terms like $-|Y|\overline{\psi}\psi\overline{\psi}\psi$ would become more relevant than terms like $m\overline{\psi}\psi$ with the result of a larger attraction and thus an even larger increment of the density of the distribution; conversely if the external field forces to decrease the density of the distribution, then terms like $-|Y|\overline{\psi}\psi\overline{\psi}\psi$ would become less relevant than terms like $m\overline{\psi}\psi$ with the result of a larger tendency to spread and hence an even larger decrement of the density of the distribution. To impose a condition of equilibrium that is to be any stable, then the term that is the most relevant at large densities must provide an effective repulsive potential: thus the constant $Y$ must be taken to have a positive value. In \cite{Fabbri:2012zc} this is what was assumed and similarly here this condition will be taken too.

This situation is also natural because the torsionally induced spin contact interaction, being connected to the presence of internal rotational degrees of freedom, is such that its resulting non-linear potential should behave as a centrifugal barrier, and so repulsively. The problem is that if this effective force is repulsive, because also the mass term gives rise to a spreading of the matter field, then neither can provide a mechanism of confinement within the matter field equations; on the other hand, if the constant is positive then the energy will always remain positive, the gravitational field attractive and singularities will form more easily: this was essentially the result of the analysis carried on in \cite{k}. But even though the condition for the formation of singularities in the gravitational sector and the condition for the equilibrium within the matter field equations seem to be worsened in each of their own domains of validity, nonetheless two effects that are bad separately may cancel one another if they are considered altogether: gravitational singularities do not form because of the repulsive forces dominating at large densities of matter \cite{Fabbri:2011mg}, the matter spreading will be confined by surrounding gravity. As a consequence, it is possible that a condition of equilibrium will arise and that it will be stable under external perturbations.

The only remaining point that need be settled about the constant $Y$ is its value: in \cite{Fabbri:2014vda} it has been thoroughly discussed how this constant does not have a value that is fixed experimentally and therefore in general it must be considered undetermined, and therefore here we will also keep its value unspecified throughout calculations, seeing what it should be in the end. So we will follow the above paper also regarding the general method of calculation.

We now have to render these arguments quantitative.

To begin we may start by noticing that with some Fierz identities it is possible to rearrange the field equations as
\begin{eqnarray}
\nonumber
&i\boldsymbol{\gamma}^{\alpha}[\partial_{\alpha}\psi
\!+\!\frac{1}{2}\partial_{\mu}(|g|^{\frac{1}{2}}e^{\mu}_{k})
|g|^{-\frac{1}{2}}e^{k}_{\alpha}\psi\!+\!iqA_{\alpha}\psi]\!-\!m\psi-\\
&-\frac{i}{8}\!\!\left[\!\frac{4}{3}Yi\overline{\psi}\{\boldsymbol{\sigma}_{\mu\rho},\!
\boldsymbol{\gamma}^{k}\}\psi\!-\!(\partial e^{k})_{\mu\rho}\right]\!\!
\{\boldsymbol{\sigma}^{\mu\rho}\!,\!\boldsymbol{\gamma}_{k}\}\psi\!=\!0
\end{eqnarray}
being the explicit expression of field equations (\ref{matter}).

With equation (\ref{matter}) written in this form the second line is the covariant force so that the constraint
\begin{eqnarray}
&\frac{4}{3}Yi\overline{\psi}\{\boldsymbol{\sigma}_{\mu\rho},
\boldsymbol{\gamma}^{k}\}\psi\!=\!(\partial e^{k})_{\mu\rho}
\label{relativequilibrium}
\end{eqnarray}
is the necessary condition for the equilibrium of the distribution described by the field and it can be read as the fact that the gravitational potential has a curl that is to be balanced by the spin density vector of the field itself.

Finding solutions to this problem requires solving gravitational field equations exactly, which are highly non-linear, while a non-linearity of less impact could be encountered in the low-speed limit, in which we may employ the approximated field equations given by (\ref{gravity}-\ref{electrostatic}) and (\ref{mass}).

Field equations (\ref{gravity}) and (\ref{electrostatic}) now can be integrated as
\begin{eqnarray}
4\!\!\!\int_{S}\!\!\!\mathrm{grad}u\!\cdot\!dS
\!\!=\!\!\!\int_{V}\!\!\!|\mathrm{grad}w|^{2}dV
\!\!+\!2Y\!\!\!\int_{V}\!\!\!|\phi^{2}|^{2}dV
\!\!+\!m\!\!\int_{V}\!\!\!\phi^{2}dV
\label{u}\\
\int_{S}\!\!\mathrm{grad}w\!\cdot\!dS\!=\!-q\!\!\int_{V}\!\!\phi^{2}dV
\label{w}
\end{eqnarray} 
so that in terms of another integration we would get the two energies $U$ and $W$ to plug in equation (\ref{mass}).

In the field equation (\ref{mass}) it is a customary analysis to require that the potential is minimum when its derivative vanishes and there if the second derivative is positive
\begin{eqnarray}
&\frac{d}{dV}\left(\frac{Y^{2}}{2m}|\phi^{2}|^{2}\!\!+\!\!Y\!\phi^{2}\!+\!U\!\!+\!W\right)\!=\!0
\label{equilibrium}\\
&\frac{d}{dV}\frac{d}{dV}\left(\frac{Y^{2}}{2m}|\phi^{2}|^{2}
\!\!+\!\!Y\!\phi^{2}\!+\!U\!\!+\!W\right)\!>\!0
\label{stability}
\end{eqnarray}
as conditions for stable equilibrium of the distribution.

In the region where these conditions hold the potential at its minimum is almost constant and negative and so
\begin{eqnarray}
&\vec{\boldsymbol{\nabla}}\!\cdot\!\vec{\boldsymbol{\nabla}}\phi\!-\!K^{2}\phi\!\approx\!0
\label{equation}
\end{eqnarray}
in which the lack of information about the distribution in the inner region forbids to know the total potential and the constant $K$ is positive but unspecified otherwise.

In general, these integrals require the knowledge of the dimension of the volume and the shape of its boundary surface, and in turn this means the knowledge of the matter field distribution in the inner region, which can only be obtained if solutions are already known, so that again some non-linearities make exact solutions impossible by construction, although we may dispose of such non-linear properties by employing further simplifying assumptions.

A specific assumption is considered in the following.
\subsection{Partially-Superposed Chiral Projections}
To get the possibility of finding solutions we must assign some extension and shape to the region occupied by the field so to guess the volume and surface of integration.

Some reasonable guess can be obtained by taking into account the results in \cite{Fabbri:2014iya}: in this reference, it was shown that the correction to the magnetic moment of charged massive fermions can be obtained as the result of the electrodynamic interaction between left-handed and right-handed semi-spinorial components at the average distance of the Compton length; as a consequence, we may try to assume that the matter field decomposed in its two chiral projections be taken to have a mean separation of the order of magnitude nearing the Compton scale.

In that paper we used the chiral representation while here we are employing the standard representation, where the difference between the left-handed and right-handed semi-spinorial components yields the small-valued semi-spinorial component while the sum of the left-handed and right-handed semi-spinorial components yields the large-valued semi-spinorial component; in the non-relativistic limit the expression of the two chiral projections tends to coincide, so that the small-valued semi-spinorial component tends to vanish and the remaining large-valued semi-spinorial component is the field $\phi$ above.

Thus we think at the field $\phi$ as the sum of the two chiral projections confined inside two compact regions split in internal part and its boundary and whose separation in average is approximately the Compton scale.

If we are in the spherically symmetric case the volume of integration becomes a radial integration while the surface of integration becomes trivial: equations (\ref{u}-\ref{w}) are
\begin{eqnarray}
4r^{2}\frac{du}{dr}\!=\!\!\!\int_{0}^{r}\!\left|\xi\frac{dw}{d\xi}\right|^{2}\!\!\!d\xi
\!+\!2Y\!\!\!\int_{0}^{r}\!\!\!\!\!|\xi\phi^{2}|^{2}d\xi
\!+\!m\!\!\int_{0}^{r}\!\!\!\!\phi^{2}\xi^{2}d\xi\\
r^{2}\frac{dw}{dr}\!=\!-q\!\!\int_{0}^{r}\!\!\!\!\phi^{2}\xi^{2}d\xi
\end{eqnarray}
as a straightforward change of coordinates shows.

The knowledge of the distribution on the boundary will allow us to obtain $U$ and $W$ in (\ref{equilibrium}-\ref{stability}), but as it is usual in general we may neglect the first and third terms of the first expression compared to the second term of the first expression and to the second expression, leaving only
\begin{eqnarray}
\frac{d}{dr}(mu\!+\!qw)\!=\!\frac{Ym}{2r^{2}}\!\!\int_{0}^{r}\!\!\!\!|\xi\phi^{2}|^{2}d\xi
\!-\!\frac{q^{2}}{r^{2}}\!\!\int_{0}^{r}\!\!\!\!\phi^{2}\xi^{2}d\xi
\label{force}
\end{eqnarray}
between which no comparison can be done at present.

In these coordinates the inner region is from the origin to a certain radius whereas the boundary is from this radius to $r$ and within the last portion we know that the distribution is given by solving (\ref{equation}) and therefore as
\begin{eqnarray}
\phi\!\approx\!\frac{Ae^{-Kr}}{r}
\label{solution}
\end{eqnarray}
in which we cannot know what is the normalization condition and the constant $A$ cannot be specified at all.

Nevertheless it is now possible to perform the integrals getting information about the behaviour at the interface between the two chiral projections of the matter.

First of all we have to require that in (\ref{force}) it must be
\begin{eqnarray}
Ym\!\!\int_{0}^{r}\!\!\!\!|\xi\phi^{2}|^{2}d\xi
\!\gg\!2q^{2}\!\!\int_{0}^{r}\!\!\!\!\phi^{2}\xi^{2}d\xi
\end{eqnarray}
encrypting the circumstance for which the gravitational attraction has to be more relevant than the internal electrodynamic repulsion: as $q^{2}\!\approx\!4\pi\alpha$ where $\alpha$ is the known fine-structure constant then it is clear that an acceptable way we have to satisfy the approximation above is
\begin{eqnarray}
Ym\!\!\int_{0}^{r}\!\!\!\!|\xi\phi^{2}|^{2}d\xi
\!\approx\!16\pi^{2}\!\!\int_{0}^{r}\!\!\!\!\phi^{2}\xi^{2}d\xi
\label{condition}
\end{eqnarray}
which will also be useful to simplify calculations.

In accordance to this assumption and (\ref{solution}) we have that
\begin{eqnarray}
\frac{Y^{2}\!A^{2}}{16\pi^{3}mr^{2}}\frac{d}{dr}\frac{e^{-2Kr}}{r^{2}}
\!+\!\!\int_{0}^{r}\!\!\!\!\eta^{-2}\!\!\!\int_{0}^{\eta}\!\!\phi^{2}\xi^{2}d\xi d\eta
\!\approx\!0
\end{eqnarray}
which is to be solved for $R$ as radial coordinate at which we have equilibrium and there we also have that
\begin{eqnarray}
\frac{d}{dr}\!\left(\!\frac{Y^{2}\!A^{2}}{16\pi^{3}mr^{2}}\frac{d}{dr}\frac{e^{-2Kr}}{r^{2}}
\!+\!\!\int_{0}^{r}\!\!\!\!\eta^{-2}\!\!\!\int_{0}^{\eta}\!\!\phi^{2}\xi^{2}d\xi d\eta\! \right)\!\!>\!0
\end{eqnarray}
for the value $R$ of the radial coordinate we found above as the radial coordinate at which the equilibrium is stable.

In the inner region there may be no information about the form of the solution but we know that regular solutions in the origin vanish and on the boundary the solution is given by $r\phi(r)\!=\!A\varphi(K\!r)$ so that, despite the fact that the remaining integrals are not elementary, they can formally be written as the primitive on the boundary minus the primitive at the origin, and because the former is known and the latter vanishes, simplifications will occur.

In fact in such a case the condition of equilibrium is
\begin{eqnarray}
\frac{Y^{2}}{8\pi^{3}mR^{5}}
\!-\!\frac{e^{2K\!R}}{K\!R\!+\!1}\lim_{\zeta\to K\!R}\int\!\!\zeta^{-2}\!\left|\int\!\!|\varphi(\zeta)|^{2}d\zeta\right|\!d\zeta\!\approx\!0
\end{eqnarray}
where the second term is a positive function of $K\!R$ only.

The condition of stability eventually reduces to the
\begin{eqnarray}
\frac{Y^{2}}{8\pi^{3}mR^{5}}
\!+\!\frac{e^{2K\!R}\,(K\!R)^{-1}}{2K^{2}\!R^{2}\!+\!6K\!R\!+\!5}
\lim_{\zeta\to K\!R}\int\!\!|\varphi(\zeta)|^{2}d\zeta\!>\!0
\end{eqnarray}
again with the second term positive function of $K\!R$ solely.

In general $K\!R$ represents the product between the depth and the width of the potential well in which the distribution is localized and for typical problems of quantum mechanics such a product is of the order of the unity so that it is reasonable to assume it is the same here.

It is straightforward to see that when this hypothesis is implemented then the two conditions above are simplified and as a consequence we obtain the condition
\begin{eqnarray}
Y^{2}\!\approx\!8\pi^{3}mR^{5}C^{2}
\end{eqnarray}
in terms of the constant $C$ being a fraction of unity but unspecified otherwise and where such a condition yields the expression of the radius at which there is equilibrium and for which this equilibrium has stability ensured.

The equivalent form $Y\!\approx\!4\pi^{2}R^{2}C$ holds if we require that such a radius be close to the Compton length.

Because the constant $Y$ has the dimension of a length squared then its squared root is interpreted as the scale at which the non-linear potentials start to become the most relevant contribution and so the equilibrium is stable where the non-linear potentials are dominant.
\subsection{Identical Particles}
In this section we will consider the particle as a whole and we will study the system constituted by two of them.

The advantage we have in considering the particle in its entirety is that it becomes possible to think at it as fully localized inside a given region of space, with two consequences: the first is that it becomes possible to implement a condition of normalization of the field density over the occupied volume, and the second is that because of this normalization condition it follows that having a precise information about the exact distribution in the inner region becomes less relevant for the knowledge of what happens on the boundary and so we are free to make some reasonable guess about the form of the density field.

In fact, we may even be free to assume that the density field is constant, whose constant value is given in terms of the normalization condition, and as a dimensional analysis would show such a constant value must be the inverse of the volume of the region: this can also be justified by thinking that an increase in the field density is determined by the reduction of the occupied volume and so that we may write $\phi^{2}\!=\!L^{-3}$ where $L$ is the radius of the region assumed to be spherical; for unidimensional cases we may write $\phi^{2}\!=\!L^{-1}\!R^{-2}$ where $L$ is the length of the region and $R$ is a typical length unspecified for now.

Now considering the effects of the external fields we have that (\ref{u}) and (\ref{w}) can be integrated giving
\begin{eqnarray}
-\mathrm{grad}(mu\!+\!qw)\!=\!-\frac{mY}{8\pi r^{2}}\langle\phi^{2}\rangle
\!+\!\frac{q^{2}}{4\pi r^{2}}
\end{eqnarray}
and employing the assumption that in this case the density scales as the inverse of the volume then
\begin{eqnarray}
-\mathrm{grad}(mu\!+\!qw)\!=\!-\frac{mY\!F}{8\pi r^{5}}\!+\!\frac{q^{2}}{4\pi r^{2}}
\end{eqnarray}
for a constant $F$ unspecified: non-linear potentials in the gravitational sector give rise to forces of gravity that are attractive and scale as $r^{-5}$ while the electrostatic forces are repulsive and scale as $r^{-2}$ as it should be expected.

For a radius much larger than the Compton length the non-linear potentials drop to zero much faster than the electrostatic forces vanish, proving that the non-linear potentials may be relevant up to the Compton scale and beyond this limit they tend to disappear very quickly.

Thus even if the non-linear potentials can empower the gravitational effects for the dynamics of particles for larger scales such effects are lost rather rapidly.

Nevertheless, the non-linear potentials may still have effects in terms of the effective forces of repulsion between two particles that are pushed onto one another.

To study such a situation we consider the Schr\"{o}dinger equation taking into account only the non-linear potentials with the highest-density term according to
\begin{eqnarray}
&\frac{1}{2m}\!\vec{\boldsymbol{\nabla}}\!\cdot\!\vec{\boldsymbol{\nabla}}\phi
\!-\!\frac{Y^{2}}{2m}|\phi^{\dagger}\!\phi|^{2}\phi\!+\!\!E\phi\!=\!0
\end{eqnarray}
and since in this case the density scales as the inverse of a length for the system then the energy levels are
\begin{eqnarray}
E\!=\!\frac{Y^{2}}{2m}\frac{1}{R^{4}}\frac{1}{L^{2}}
\end{eqnarray}
as it can be checked; it is evident that the energy levels have exactly the dependence of the Fermi energy levels.

Also if $Y\!=\!\pi R^{2}$ then the non-linear potential energy levels and the Fermi energy levels coincide exactly.

For such a system the Hamiltonian is given by
\begin{eqnarray}
&\boldsymbol{H}\!=\!\frac{Y^{2}}{2m}\phi^{\dagger}\phi\,\phi^{\dagger}\phi
\end{eqnarray}
and identities $\phi^{\dagger}\phi\,\phi^{\dagger}\phi\!\equiv\!
\phi^{\dagger}\vec{\boldsymbol{\sigma}}\phi\!\cdot\!\phi^{\dagger}\vec{\boldsymbol{\sigma}}\phi$ where $\vec{\boldsymbol{\sigma}}$ are the Pauli matrices allow the Hamiltonian to be written as
\begin{eqnarray}
&\boldsymbol{H}\!=\!\frac{Y^{2}}{2m}\phi^{\dagger}\vec{\boldsymbol{\sigma}}\phi
\!\cdot\!\phi^{\dagger}\vec{\boldsymbol{\sigma}}\phi
\end{eqnarray}
for the total potential of the interacting system.

The part that describes the mutual interaction is
\begin{eqnarray}
\boldsymbol{H}_{12}\!=\!\frac{\pi}{L_{12}}\vec{\boldsymbol{s}}_{1}\!\cdot\!\vec{\boldsymbol{s}}_{2}
\end{eqnarray}
as the scalar product of the two spins being attractive for anti-parallel spins and repulsive for parallel spins as the exchange force described by the Heisenberg-Dirac model.

The non-linear potentials and the Pauli exclusion have effects that at least in one dimension are analogous.

And if parameters $R$ in these two sub-sections coincide then also their magnitude is similar indeed.
\section{Spin States}
We consider again two particles; for such a system we study the algebraic properties of non-linear potentials.

In general the spinor can be decomposed in terms of its left-handed and right-handed semi-spinorial components which are themselves decomposable in spin-up and spin-down parts, so that in chiral representation we may write 
\begin{eqnarray}
&\psi\!=\!\left(\begin{tabular}{c}
$\psi_{L}$\\
$\psi_{R}$
\end{tabular}\right)\!=\!\left(\begin{tabular}{c}
$\psi_{L}^{+}$\\
$\psi_{L}^{-}$\\
$\psi_{R}^{+}$\\
$\psi_{R}^{-}$
\end{tabular}\right)
\end{eqnarray}
while the passage from this to the standard representation is given in terms of the following linear combination
\begin{eqnarray}
&\psi\!=\!\frac{1}{\sqrt{2}}\left(\begin{tabular}{c}
$\psi_{R}\!+\!\psi_{L}$\\
$\psi_{R}\!-\!\psi_{L}$
\end{tabular}\right)\!=\!\frac{1}{\sqrt{2}}\left(\begin{tabular}{c}
$\psi_{R}^{+}\!+\!\psi_{L}^{+}$\\
$\psi_{R}^{-}\!+\!\psi_{L}^{-}$\\
$\psi_{R}^{+}\!-\!\psi_{L}^{+}$\\
$\psi_{R}^{-}\!-\!\psi_{L}^{-}$
\end{tabular}\right)
\end{eqnarray}
as it is a simple unitary rotation; the standard representation is useful in non-relativistic limits because in this case there remains only one semi-spinorial component which is itself split in spin-up and spin-down parts as
\begin{eqnarray}
&\psi\!\approx\!\left(\!\begin{tabular}{c}
$\phi$\\
$0$
\end{tabular}\!\right)\!=\!\left(\!\begin{tabular}{c}
$\phi^{+}$\\
$\phi^{-}$\\
$0$\\
$0$
\end{tabular}\!\right)
\end{eqnarray}
by construction: combining these two expressions yields
\begin{eqnarray}
&\phi^{\pm}\!\approx\!\frac{1}{\sqrt{2}}\left(\psi_{R}^{\pm}\!+\!\psi_{L}^{\pm}\right)\\
&\psi_{R}^{\pm}\!\approx\!\psi_{L}^{\pm}
\end{eqnarray}
as the relationship between the two representations when the non-relativistic approximation is considered.
\subsection{Degeneracies}
Now we may turn our attention to the spin algebra.

The last obtained relationship between the two representation shows that in the passage from chiral to standard representation the chiral projections are mixed, but the spin algebra is respected, or in equivalent words, in going from the relativistic case, in which chirality is well defined, to the non-relativistic regime, where we have a single field given as the superposition of the two chiral projections, we have that spin-up/down components are summed together to give spin-up/down components.

What this means is that in the non-relativistic field given in terms of the superposition of the two chiral components we have used above we may not know what is the spin of the non-relativistic field but in either case we know that it is the sum of two chiral projections having the same spin; thus we are in the situation of superposition of two fields with two aligned spins and this situation is what ensures the effects of the Pauli exclusion.

If we had a system constituted by two fields that were totally independent then they might have aligned spins but also anti-aligned spins, and we would not have an ensured action of the repelling forces: by taking the spin direction as the $\mathrm{z}$-axis 
$\boldsymbol{\sigma}_{\mathrm{z}}\phi_{\pm}\!=\!\pm\phi_{\pm}$ and so in the case of two fields with aligned spins there is no information that can be obtained but in the case of two fields with anti-aligned spins we get the general rule for the spin sum 
\begin{eqnarray}
\nonumber
&\phi_{\pm}^{\dagger}\phi_{\mp}\!=\!
\phi_{\pm}^{\dagger}\boldsymbol{\sigma}_{\mathrm{z}}\boldsymbol{\sigma}_{\mathrm{z}}\phi_{\mp}
\!=\!-\phi_{\pm}^{\dagger}\phi_{\mp}
\end{eqnarray}
showing that $\phi_{\pm}^{\dagger}\phi_{\mp}\!\equiv\!0$ and therefore we have that if from the non-linear potential we discard the self-interaction extracting only the part that describes the mutual interaction then this mutual interaction is equal to zero.

Two particles with opposite spin have the non-linear potential yielding a mutual interaction that vanishes so that they can superpose, allowing for degeneracy.

The torsionally induced spin-contact interactions and the Pauli exclusion have equal phenomenology.
\section{Comments}
We summarize now what we have done.

We have considered a first case in which the matter field distribution was taken to be the superposition of the two chiral projections: the two chiral projections are the two components of the same massive field and so they always have a partial superposition; on the other hand, the two chiral projections have by definition opposite helicities and by construction opposite momenta, so that they are in the same spin state, and therefore they are never entirely superposed. Then a first consequence is that the interaction takes place at the interface between the two chiral projections; and a second consequence is that the Compton scale is taken to be the average separation of the two components. Then we studied the dynamics.

A first issue we had was that the torsional contribution to the gravitational attraction was more relevant than the electrodynamic repulsion if (\ref{condition}) held and this condition additionally helped us in getting rid of the unknown normalization constant $A$ from calculations; furthermore, for typical problems of quantum mechanics in which the product between the depth and the width of the potential is a constant of the system we could get rid of the unknown potential $K$ from computations. As a result we had obtained that all hypotheses condensed into the single requirement $\sqrt{Y}\!\approx\!R$ with $R$ then being the length that is interpreted as the scale at which the non-linear potential is dominant and therefore the matter field distribution has an equilibrium which is also stable.

This system is constituted by the two chiral projections in partial superposition and for which the torsional contribution to the gravitational attraction and the torsional intrinsic repulsion balance giving rise to an equilibrium that is stable for the matter field distribution.

Such a matter field distribution can be thought to be a classical description of the quantum particle and further we have considered a second case in which we took two identical particles: the torsional contribution to the gravitational pull scaled as $r^{-5}$ and the electrostatic repulsion scaled as $r^{-2}$ as usual, so that for radii larger than the Compton length the gravitational pull dropped faster than the electrostatic repulsion, and the expected dynamics was recovered; also the torsionally induced spin-contact interaction manifested as the non-linear potential was rendered feeble, but still non-negligible, and eventually its effects became similar to those one would expect between two identical particles as given by the Pauli exclusion in terms of the Fermi energy and as the exchange force described by the Heisenberg-Dirac Hamiltonian.

What this suggested was that the torsional effects and the Pauli exclusion might have a common origin.

In fact, we also found that the torsional effects describing the mutual interaction between two particles with anti-aligned spin disappeared allowing for the particle degeneracy, and thus showing that the torsional effects and the Pauli exclusion have equal phenomenology.

In the dynamics that is internal to a fundamental field and which is cyphered as the interaction at the interface between the two chiral projections the presence of torsion has both the effect of increasing the energy content in the field equations for gravity and the effect of giving rise to the non-linear potentials in the field equations for matter, both making gravitation stronger than usual and adding a repulsive force; given a certain critical radius, for more extended distributions matter is subject to an external gravitational pull that keeps it confined within a compact region while for more localized distributions matter has an intrinsic barrier that prevents it to collapse into a singularity. And the equilibrium has stability.

In the dynamics between two fundamental fields gravity is negligible and also in the matter field equations torsion is weaker but still non-negligible and its residual effects are essentially similar to the Pauli exclusion.

Therefore matter field distributions may have equilibrium with stability displaying quantum features.
\section{CONCLUSION}
In this paper, we have discussed the fact that the presence of torsion has two effects, of which one is augmenting the energy content of the gravitational field equations and thus the gravitational pull and the other is giving rise to non-linear potentials in the matter field equations that are effectively repulsive; the former effect is such that it makes gravitation the dominant force acting from the exterior and thus providing the conditions needed for the material confinement inside a localized region while the latter effect is intrinsically repulsive making it impossible for matter to collapse into a point and thus forbidding singularity formation, and overall we have demonstrated that the fermionic matter distributions are in equilibrium with stability with no necessity to implement quantization prescriptions but considering classical fields solely.

Only we need not neglect torsion, although this is not a requirement since torsion is already present in the most general case for the geometry of classical field theory.

What we have found most curious in this discussion is the fact that these torsional spin effects are important insofar as the matter wave character is shown.


\begin{thebibliography}{20}
\bibitem{C1} 
E.Cartan,
\textit{Annales Sci.Ecole Norm.Sup.} \textbf{40}, 325 (1923).
\bibitem{C2} 
E.Cartan,
\textit{Annales Sci.Ecole Norm.Sup.} \textbf{41}, 1 (1924).
\bibitem{C3} 
E.Cartan,
\textit{Annales Sci.Ecole Norm.Sup.} \textbf{42}, 17 (1925).
\bibitem{C4} 
E.Cartan,
\textit{Compt.Rend.Acad.Sci.} \textbf{174}, 593 (1922).
\bibitem{S} 
D.W.Sciama, in \textit{Recent Developments in\\
General Relativity} (Pergamon, Oxford, 1962).
\bibitem{K}
T.W.B.Kibble, 
\textit{J.Math.Phys.} \textbf{2}, 212 (1961).
\bibitem{h-h-k-n}
F.W.Hehl, P.Von Der Heyde, G.D.Kerlick and\\
J.M.Nester, \textit{Rev. Mod. Phys.} \textbf{48}, 393 (1976).
\bibitem{Fabbri:2013isa}
L.Fabbri,
\textit{Int.J.Theor.Phys.} \textbf{53}, 3744 (2014).
\bibitem{Fabbri:2014zya}
L.Fabbri,
\textit{Gen.Rel.Grav.} \textbf{47}, 1837 (2015).
\bibitem{Fabbri:2014dxa} 
L.Fabbri,
\textit{Int.J.Geom.Meth.Mod.Phys.}\textbf{12},1550099(2015).
\bibitem{Magueijo:2012ug}
J.Magueijo, T.G.Zlosnik and T.W.B.Kibble,\\
\textit{Phys.Rev.D} \textbf{87}, 063504 (2013).
\bibitem{n-j--l}
Y.Nambu, G.Jona--Lasinio,
\textit{Phys. Rev.} \textbf{122}, 345 (1961).
\bibitem{Fabbri:2012zc} 
L.Fabbri,
\textit{Int.J.Mod.Phys.D}\textbf{22}, 1350071 (2013).
\bibitem{k}
G.D.Kerlick,
\textit{Phys. Rev. D} \textbf{12}, 3004 (1975).
\bibitem{Fabbri:2011mg}
L.Fabbri,
\textit{Int.J.Theor.Phys.} \textbf{52}, 634 (2013).
\bibitem{Fabbri:2014vda} 
L.Fabbri,
\textit{Mod.Phys.Lett.A}\textbf{29}, 1450133 (2014).
\bibitem{Fabbri:2014iya}
L.Fabbri,
\textit{Int.J.Theor.Phys.} \textbf{55}, 669 (2016).
\end{thebibliography}
\end{document}